\documentclass[letterpaper]{llncs}
\usepackage{mathptmx}
\usepackage{hyperref}
\usepackage{tablefootnote}
\usepackage{listings}
\usepackage{graphicx}
\usepackage{epstopdf}
\usepackage{cleveref}
\usepackage{xspace}
\usepackage{xcolor}
\usepackage{amsmath}
\usepackage{amssymb}
\usepackage{subcaption}
\usepackage{multirow}
\usepackage{algorithm}
\usepackage{algpseudocode}
\usepackage{mathtools}
\usepackage{bbm}
\usepackage{sidecap}

\makeatletter
\def\lst@makecaption{%
    \def\@captype{table}%
    \@makecaption
} 
\renewcommand{\ALG@beginalgorithmic}{\small}
\makeatother

\newcommand{\OFP}{Oakforest-PACS\xspace}

\newcommand{\Pmax}{\ensuremath{P^\text{max}}\xspace}

\newcommand{\bq}{\begin{equation}}
\newcommand{\eq}{\end{equation}}

\newcommand{\byte}{\mbox{byte}\xspace}
\newcommand{\second}{\mbox{s}\xspace}

\newcommand{\flop}{\mbox{flop}\xspace}

\newcommand{\GBS}{\mbox{G\byte/\second}\xspace}

\newcommand{\TFS}{\mbox{T\flop/\second}\xspace}
\newcommand{\PFS}{\mbox{P\flop/\second}\xspace}

\newcommand{\MHZ}{\mbox{MHz}\xspace}

\newcommand{\GiB}{\mbox{GiB}\xspace}
\newcommand{\MiB}{\mbox{MiB}\xspace}
\newcommand{\KiB}{\mbox{KiB}\xspace}

\algnewcommand{\algorithmicgoto}{\textbf{go to}}%
\algnewcommand{\Goto}[1]{\algorithmicgoto~\ref{#1}}%

\begin{document}
%\title{Chebyshev Filter Diagonalization on Compute Nodes with Transparent Hierarchical Memory Access}
\title{Chebyshev Filter Diagonalization on Modern Manycore Processors and GPGPUs}
\author{Moritz Kreutzer\inst{1} \and Georg Hager\inst{1} \and Dominik
  Ernst\inst{1} \and Holger Fehske\inst{2}\and\\ Alan
  R. Bishop\inst{3} \and Gerhard Wellein\inst{1}} \institute{Erlangen
  Regional Computing Center (RRZE), Friedrich-Alexander University of
  Erlangen-Nuremberg \and Institut f\"ur Physik,
  Ernst-Moritz-Arndt-Universit\"at Greifswald \and Theory, Simulation
  and Computation Directorate, Los Alamos National Laboratory }
\maketitle
\begin{abstract}

  Chebyshev filter diagonalization is well established in
  quantum chemistry and quantum physics to compute bulks of
  eigenvalues of large sparse matrices. Choosing a block vector
  implementation, we investigate optimization
  opportunities on the new class of high-performance compute devices
  featuring both high-bandwidth and low-bandwidth memory. We focus on
  the transparent access to the full address space supported by both
  architectures under consideration: Intel Xeon Phi ``Knights
  Landing'' and Nvidia ``Pascal.''
 
  We propose two optimizations: (1) Subspace blocking is applied for
  improved performance and data access efficiency.  We also show that
  it allows transparently handling problems much larger than the
  high-bandwidth memory without significant performance penalties. (2)
  Pipelining of communication and computation phases of successive
  subspaces is implemented to hide communication costs without extra
  memory traffic.
  
  As an application scenario we use filter diagonalization studies
  on topological insulator materials.
  Performance numbers on up to 512 nodes of the \OFP and Piz Daint
  supercomputers are presented, achieving beyond 100\,\TFS
  for computing $10^2$ inner eigenvalues of sparse matrices of
  dimension $10^9$.
  
\end{abstract}

%
% - device blocking weg (3.5 kuerzen - nur 1 subspace blocking)
% - Alg. 4 checken, Fig. 3 aendern
% - 
%
\section{Introduction and related work}
Stacked memory technologies such as HBM2 and MCDRAM have boosted the attainable main memory bandwidth by a factor of five to six compared to conventional multicore  systems. Soon after the commercial availability of these technologies, three out of the ten most powerful supercomputers were equipped with the new fast memories (see the TOP500~\cite{TOP500} list as of June 2017). Typically holding 16\,\GiB of data, the size of stacked memories is still very limited and hierarchical concepts have been implemented, offering additional large DDR4 memory spaces.  The two major players as of today, Intel with its ``self-hosted'' Xeon Phi ``Knights Landing'' (KNL) series and Nvidia with its ``Pascal'' (P100) GPGPUs, implement these hierarchical concepts in different ways. While the KNL is directly connected to the DDR4 partition, the P100 accesses the large host node memory through the PCIe interface. However, both architectures are capable of transparently addressing the complete (slow and large) memory on a node, thereby offering easy access to large data sets.

The computation of bulks of eigenvalues of large sparse matrices is very data intensive, both in terms of bandwidth demands (i.e., low computational intensity) and data set sizes. Subspace projection using polynomial filters based on the Chebyshev iteration is an efficient approach for the computation of extremal and interior eigenvalues in quantum physics and quantum chemistry. Application areas include inner eigenvalue problems in the context of modeling graphene or topological insulator materials~\cite{SF12,SFFV12} or the solution of an eigenvalue problem in density functional theory~\cite{ZHOU2006172}. Beyond eigenvalue computations, Chebyshev polynomials can be used as acceleration techniques for linear solvers (see, e.g.,~\cite{Saad84,BASERMANN1997}) in various application areas (e.g., power flow modeling~\cite{Kamiabad2013,LI201487,LiLi2016}). Moreover, the closely related kernel polynomial method (KPM) (see~\cite{KPMRMP} for a review on KPM and its relation to Chebyshev polynomials) also relies on evaluating those polynomials to calculate spectral properties of sparse matrices, such as the density of states~\cite{BIBC13,Di_Napoli2016:838}.

From a computational perspective, the evaluation of Chebyshev polynomials is a simple series of vector operations and sparse matrix-vector multiplications (SpMV). It allows for kernel fusion to increase the computational intensity~\cite{Kreutzer15KPM}. Global communication can be avoided or limited to a single invocation for the full-degree polynomial. In the above application scenarios the polynomial is usually evaluated for multiple vectors and the algorithm can be reformulated to use blocks of vectors. This further increases the computational intensity and pushes the corresponding sparse matrix-multiple-vector multiplication (SpMMV) towards regular data access~\cite{Kreutzer15KPM}. We emphasize that the benefits of SpMMV have been known for a long time~\cite{Gropp99} but have only recently  gained renewed interest (see, e.g.,~\cite{Liu12,Aktulga14,Anzt2015}).

Performance modeling, code optimization strategies, and parallel scalability
studies have been presented
%by the authors
for KPM~\cite{Kreutzer15KPM} and Chebyshev filter diagonalization~\cite{Pieper15ChebFD}. These investigations were performed on two \PFS-class supercomputers: the SuperMUC-Phase2 system\footnote{\url{https://www.lrz.de/services/compute/supermuc/systemdescription/}}, which is based on the Intel Xeon Haswell, and the first phase of the Piz Daint supercomputer\footnote{\url{http://www.cscs.ch/computers/piz_daint}} (Cray XC30), using Intel Xeon Sandy Bridge processors and Nvidia K20 GPGPUs.

\subsection{Contribution}
This paper extends existing work towards the new class of supercomputers using compute nodes that feature both high- and low-bandwidth memory and transparent access to the full memory address space of a node. The systems under consideration are phase two of Piz Daint and the \OFP\footnote{\url{http://www.cc.u-tokyo.ac.jp/system/ofp/index-e.html}} system, representing the Nvidia P100-based accelerator and the standalone Intel Xeon Phi approach, respectively. As of June 2017 these supercomputers were ranked on position 3 and 7 of the TOP500 list. 

We first investigate the attainable bandwidth within the compute nodes, focusing on the usage modes for accessing the low-bandwidth partitions. Concerning the transparent use of the low-bandwidth memory, the tighter hardware integration allows much faster access to large data sets on the KNL. Then we report on efforts porting and optimizing the code for the new compute device architectures and analyze the attainable performance levels and hardware bottlenecks if the working set data fits into the high-bandwidth memory. Our block vector implementation (i.e., storing all $n_s$ vectors in a consecutive array)  and the simplicity of the algorithm allow for a straightforward implementation of subspace blocking strategies. We perform these in three directions: (1) We block for optimal compute performance, i.e., the computation of the Chebyshev filter polynomial is restricted to a subset of $n_b$ vectors at a time. (2) We show that the subspace blocking is adequate to enable the efficient use of transparent DRAM data access for large problems. (3) We interchange the original order of polynomial evaluation in combination with a pipeline strategy and demonstrate that overlapping of communication and computation between successive subblocks of size $n_b$ can be realized, avoiding the redundant memory transfers of standard communication hiding mechanisms in SpMMV. We investigate these approaches using scalable test cases (sparse matrices) from eigenvalue computations for topological insulator simulations together with realistic parameter settings for the filter diagonalization algorithm. We also show that these kinds of computations fit very well to this new class of supercomputers. 

As our library is available as open-source software, our implementations and approaches can be easily adapted by the large community using numerical methods that involve the evaluation of Chebyshev polynomials of large sparse matrices.

\subsection{Hardware Testbed} \label{sect:Hardware}

\begin{table}[t]
    \centering
    \begin{tabular}{llccc}
        & & \textbf{P100} & & \textbf{KNL} \\
        \hline
        Vendor & & Nvidia & & Intel \\
        Model & & Tesla P100 & & Xeon Phi 7250 \\
        Codename & & Pascal & & Knights Landing \\
        Cores & & 1792 (FP64 CUDA cores) & & 68 (64 used) \\
        Clock frequency & [\MHZ{}] & 1328--1480 & & 1400 \\
        Peak performance & [\TFS{}] & 4.7--5.3 & & 3 \\
        L2 cache capacity & [\MiB{}] & 4 & & 34 \\
        Fast memory technology & & HBM2 & & MCDRAM \\
        Fast memory capacity & [\GiB{}] & 16 & & 16 \\
        Slow memory capacity & [\GiB{}] & 64 & & 96 \\
    \end{tabular}\smallskip
    \caption{Key architectural features of the two compute
      devices. The slow memory partition uses DDR4 memory technology on
      both systems.}
    \label{tab:arch}
\end{table}
The two supercomputers considered in the present work harness non-standard
compute devices to bring forth their massive computational power. The Piz
Daint system consists of 5,320 nodes, each equipped with an Intel Xeon
E5-2690v3 compute node hosting one Nvidia Tesla ``Pascal''
(P100) GPGPU. \OFP features 8,208 compute nodes, each with a self-hosted
Intel Xeon Phi 7250 ``Knights Landing'' manycore CPU.
In \Cref{tab:arch} we summarize the key features of the
P100 and the KNL.
From a high level point of view, both architectures have similar memory
organization and key performance figures. However, technical
implementations (e.g., SIMD vs SIMT execution; slow memory organization)
and programming approaches (e.g., access to slow memory) are
substantially different. As we focus in this work on large data sets
and ways to use the slow memory, a more extensive evaluation of the
different memory modes and the respective attainable data access rates
is provided in the next section. Finally, the network structure of both
supercomputers is briefly discussed in \Cref{sect:large-scale}.

\subsubsection{Memory subsystems and operating modes}

A crucial difference between both architectures is their basic operating mode.
KNL is self-hosted, i.e., everything, including the operating system and
management processes, runs on the compute device. The processor features a large
partition of slow DDR4 memory and a small partition of fast MCDRAM memory.
It can be configured such that each of them is visible to the programmer as a separate ccNUMA
domain (``flat mode''). If both domains should be used, the programmer explicitly needs to specify the data location and, if required, copy data between the domains.
%The preferred mode for applications whose data set fits into
%the fast memory is to only use the respective NUMA domain. 
Another operating mode uses MCDRAM as a transparent cache for the DDR4 memory
(``cache mode''). In this case all memory requests go to the MCDRAM; if data is not available there it will be loaded from DDR4 memory transparently to the MCDRAM and delivered to the processing units. No explicit data management is required by the programmer.
%This
%is useful for data sets exceeding the MCDRAM capacity. 

The P100 GPGPU is installed as an accelerator via PCI-Express. The device itself only contains the fast HBM2 memory. In case the data sets exceed its capacity, the host memory has to
be used. This can be done via explicit CUDA calls that copy data between host and GPGPU. The Pascal architecture is the first to support a transparent view to device memory and full host memory.  Similar to the cache mode on KNL, this ``Unified Memory''  feature  enables transparent data transfers
between the host and the device (``managed mode'').  Programmers need to allocate data with a special function
(\texttt{cudaMallocManaged()}).  Data transfers between host and GPGPU is then managed automatically by the Page Migration Engine (PME).

\begin{figure}[tb]
\centering
\includegraphics*[width=.75\linewidth]{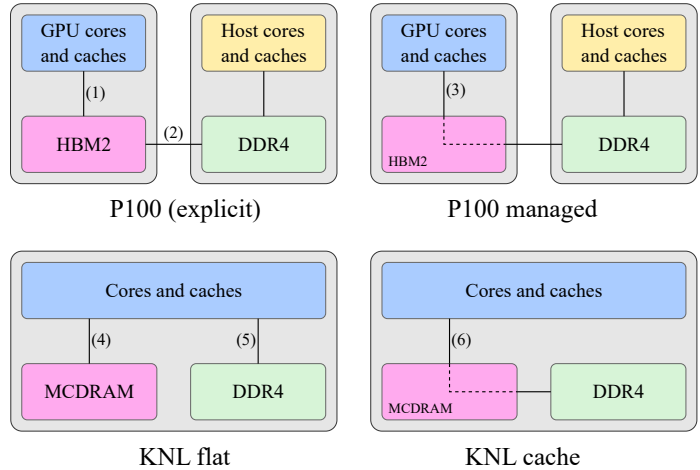}
\caption{Memory access modes on P100 (top) and KNL (bottom): Explicit management (left) of the full memory vs.\ transparent access (right) to the full memory.}
\label{fig:mem}
\end{figure}

The memory subsystems and operating modes are illustrated in \Cref{fig:mem}. 
For data sets fitting into the fast memory partitions, the operating modes on the
left are preferred and data should be transferred via path (1) or (4). If the
data set exceeds 16\,\GiB, KNL offers the cache mode which corresponds to path (6).
Explicit transfers via path (1)+(2) can be used on P100, but require explicit coding of the data transfers by the programmer. 
This can be avoided using the managed mode, i.e., data path (3), which provides a transparent view of the complete address space of the host and the GPGPU.

In order to get estimates for achievable performance we investigate attainable bandwidth numbers for accessing large consecutive data sets, which is the typical memory access scenario for the application considered in this work. We use the STREAM benchmark~\cite{McCalpin95} and adapt it to the different memory access modes. Appropriate data set sizes are chosen to measure the different bandwidth paths shown in \Cref{fig:mem}, i.e., for transparent access to the slow memory we use data sets larger than the fast memories. The measurements for all relevant data path combinations are shown in \Cref{tab:mem}.
\begin{table}[b]
\centering
\begin{tabular}{llcccc}
    & \textbf{Mode} & \textbf{Copy} & \textbf{Scale} & \textbf{Add} &
    \textbf{Triad} \\
    \hline
    \parbox[t]{4mm}{\multirow{3}{*}{\rotatebox[origin=c]{90}{P100}}} 
    & HBM2 (1) & 542 & 542 & 556 & 557 \\
    & DDR4-HBM2 explicit (1)+(2) & 13 & 13 & 12 & 12 \\
    & DDR4-HBM2 managed (3) & 3 & 2 & 3 & 3 \\
    \hline
    \parbox[t]{4mm}{\multirow{3}{*}{\rotatebox[origin=c]{90}{KNL}}} 
    & MCDRAM (4) & 466 & 468 & 481 & 489 \\
    & DDR4 (5) & 81 & 81 & 85 & 85 \\
    & DDR4-MCDRAM cache (6) & 60 & 60 & 60 & 59 \\
\end{tabular}\smallskip
\caption{Memory bandwidth in \GBS for different operating modes as illustrated in
\Cref{fig:mem} using the STREAM benchmark. On the P100 pinned memory was used for the explicit data access (``DDR4-HBM2 explicit'').}
\label{tab:mem}
\end{table} 
On both architectures the highest bandwidth is naturally attained when using the fast memory only. Access speed to the slow memory component is substantially higher on the KNL owing to its on-chip DDR4 memory controllers, while host memory access on P100 is limited by the capabilities of the PCIe 3.0$\times$16 interfaces. For explicit slow memory access approximately 75\% of the maximum uni-directional PCIe bandwidth can be attained on the P100. However, the bandwidth for transparent access (``managed mode''; path (3) in \Cref{fig:mem}) breaks down to 2--3\,\GBS in our benchmarks, which may severely restrict the use of this mode in real world applications. These low transfer rates are caused by the PME, which handles all remote page faults generated by the GPGPU and tries to consolidate them into consecutive PCIe data transfers. An analysis of the ``Host to Device Transfers'' using the Nvidia profiler shows that the average transfer size granularity even for simple kernels like STREAM is in the order of 40\,\KiB, at which the  PCIe 3.0$\times$16 interface can deliver only 2--3\,\GBS. The minimum message size is 4\,\KiB (page size of host node), and the maximum size is close to 1\,\MiB (all threads executing concurrently access consecutive data). 

In summary, the transparent access to large consecutive data sets in the slow memory on the P100 is twenty to thirty times slower, while the access to the fast memory is 25\% faster than on the KNL.

%\Cref{tab:mem} summarizes attainable bandwidth numbers for data transfers via any
%of the discussed paths. Benchmarks \#1, \#2, and \#4 use vectors of length $2^{29}$, 
%benchmark \#3 has vector length $4\times2^{29}$, and \#5 and \#6 has vector
%length $5\times2^{29}$. Additionally, benchmark \#2 uses pinned memory.

\subsection{Software Testbed}

All computations were carried out using (real or complex) double precision data.
Index values are 4-\byte integers. For the P100, the CUDA toolkit 
in version 8.0.44 was used for compilation and the respective cuBLAS version was
employed as a baseline implementation. The Intel C Compiler (ICC) version 17.0.1 was 
used for KNL with the corresponding MKL and MPI versions. On Piz Daint we used
Cray MPICH 7.5.0.

The performance numbers we present for the Chebyshev
Filter Diagonalization kernel are median values from ten consecutive
runs applying the full filter
polynomial. Before the actual measurements, one additional
warmup run was performed on the assigned set of nodes. No error bars are given for the
performance results because the variations were small ($\le$ 5\%).

\section{Chebyshev Filter Diagonalization}
We investigate Chebyshev Filter Diagonalization (ChebFD) as a
representative algorithm for large-scale and efficient eigenvalue
computations. Filter diagonalization is frequently used to find a set
of inner eigenstates of a sparse matrix $H$ in a given search interval
of eigenvalues. It uses a window function approximated by a
polynomial filter of degree $n_p$ to project a subspace of $n_s$ search vectors to a given search interval of eigenvalues. A comprehensive description of this method is
given in~\cite{Pieper15ChebFD}. The computational core of
ChebFD is the application of the polynomial filter together with the
computation of the Chebyshev moments. This is shown in
\Cref{alg:chebfd_vanilla} for a formulation with block vectors.
\begin{algorithm}[tb]
\caption{Application of the ChebFD polynomial filter to block vectors.}
\begin{algorithmic}[1]  
    \State $\vec U$\ :=\ $\vec u_1,\dots,\vec u_{n_s}$\Comment{define block vector}
    \State $\vec W$\ :=\ $\vec w_1,\dots,\vec w_{n_s}$\Comment{define block vector}
    \State $\vec X$\ :=\ $\vec x_1,\dots,\vec x_{n_s}$\Comment{define block vector}
    \State $\mathrlap{\vec U}\phantom{\vec W} \gets (\alpha\vec H + \beta \mathbbm
        1) \vec X$ \Comment{\texttt{spmmv()}}
    \State $\vec W \gets 2 (\alpha\vec H + \beta \mathbbm 1) \vec U - \vec X$
    \Comment{\texttt{spmmv()}}
    \State $\mathrlap{\vec X}\phantom{\vec W} \gets g_0 c_0 \vec X + g_1 c_1
    \vec U + g_2 c_2 \vec W$ \Comment{\texttt{baxpy()+bscal()}}
    \For{ $p=3$\ to\ $n_p$ }
    \State $\mbox{swap} ( \vec W ,  \vec U )$
    \State $\vec W \gets 2 (\alpha\vec H + \beta \mathbbm 1) \vec U -  \vec W$ 
    \State $\mathrlap{\vec \eta_p}\phantom{\vec X} \gets \langle \vec W, \vec U \rangle$ 
    \State $\mathrlap{\vec \mu_p}\phantom{\vec X} \gets \langle \vec U, \vec U \rangle$ 
    \hfill{\smash{\raisebox{\dimexpr.5\normalbaselineskip+.5\jot}{$\left.\begin{array}{@{}c@{}}\\{}\\{}\\{}\\{}\end{array}\right\}\triangleright\text{\textsc{chebfd\_op($H,\vec U, \vec W,
                                    \vec X$)}}$}}}
    \State $\mathrlap{\vec X}\phantom{\vec W} \gets \vec X + g_p c_p \vec W$ 
    \EndFor
\end{algorithmic}
\label{alg:chebfd_vanilla}
\end{algorithm}
Basic numerical operations involved in the filter application kernel
(lines 7-13) are a SpMV
involving a large sparse matrix $H$ and a series of scaled vector
addition kernels (i.e., BLAS1 kernels). These kernels can be formulated
as a single SpMMV operation involving special scaling
factors and offset computations (line 9). In lines 10/11 the Chebyshev
moments are computed; they are used to monitor the number of
eigenstates in the search interval, which is not known
a priori. Finally, in line 12 the ``filtered vector'' is updated. As
the polynomial filter of degree $n_p$ is applied independently
to $n_s$ search vectors, a block formulation as indicated in
\Cref{alg:chebfd_vanilla} can be used. In particular, the block
variant of SpMV, i.e., the SpMMV kernel, is favorable in terms of computational intensity since the
matrix data ($H$) has to be loaded only $n_p$ times instead of $n_p
\times n_s $ times if the full filter polynomial is applied separately to
each vector in the SpMV. Note that the block formulation of the vector kernels
does not impact their computational efficiency as $n_s$ BLAS1 type
operations are still involved, e.g., in line 10 $n_s$ independent dot products
are computed.

The full filter diagonalization algorithm requires orthogonalization
of the $n_s$ ``filtered vectors'' in the block vector ${\vec
  X}$ after applying the filter above and before restarting the
procedure. A rank-revealing technique such as SVQB~\cite{SW02} or
TSQR~\cite{DGHL12} is used in the orthogonalization step, but as its
contribution to the overall runtime is typically small for reasonably
large filter polynomial degrees $n_p$ we do not include it in the
performance analysis.

At this point we must emphasize that performance numbers presented
in~\cite{Pieper15ChebFD} use the ChebFD formulation presented here, while
in~\cite{Kreutzer15KPM} only the Chebyshev moments have been computed,
i.e., \Cref{alg:chebfd_vanilla} without line 12.

\subsection{Physical Application and Problem Setting} \label{sect:setting}
We have chosen the computation of a bulk of central eigenstates of a
topological insulator as a test case for our performance study. Such 
applications are of current interest in quantum physics research. The
model Hamiltonian~\cite{SRAF12} describing the topological insulator acts on a
discrete 3D lattice of size $N_x \times N_y \times N_z$ carrying four
degrees of freedom per site. The matrix formulation leads to a sparse
matrix of size $n=4 \times N_x \times N_y \times N_z$ with an average
of $n_{nzr}=13$ complex double precision non-zero elements per matrix
row (denoted by ``Topi-$N_x$-$N_y$-$N_z$''). The matrices have several
subdiagonals leading to a structure similar to 3D stencil. Please
see~\cite{Kreutzer15KPM,Pieper15ChebFD} for more details on the model
Hamiltonian and its mapping to a sparse matrix. Relevant problem
parameter settings in topological insulator research are matrix
dimensions of $n=10^6,\ldots,10^8$ and search spaces of
$n=10^2,\ldots,10^3$. In terms of algorithmic efficiency it has been
shown in~\cite{Pieper15ChebFD} that high polynomial degrees ($n_p
\approx 10^3$) deliver best results.

%\begin{itemize}
%    \item Topological insulators
%    \item nnzr=13
%    \item complex double precision
%    \item $n$: matrix/vector dimension
%    \item $n_p$: polynomial degree (defined by application)
%    \item $n_s$: number of search vectors (defined by application)
%    \item $n_d \leq n_s$: number of device vectors (defined by hardware)
%    \item $n_b \leq n_d$: number of kernel vectors (defined by performance)
%    \item Parallelisierungs- und Vecktorisierungsstrategie (accross rhs vector) nicht vergessen zu erwaehnen...  
%\end{itemize}

\section{Node-level implementation and performance analysis}

The compute kernels and implementation alternatives discussed in the
following are available for download at (\url{https://bitbucket.org/essex/ghost}).
As parallelization
approaches we use OpenMP for KNL (and CPU architectures) and CUDA for
Nvidia GPGPUs. Best performance on the KNL for our application is
typically achieved with four OpenMP threads per core.

\subsection{Implementation}

The structure of the ChebFD algorithm presented
in~\Cref{alg:chebfd_vanilla} can easily be mapped to a series of
vector operations (of BLAS1 type) and a SpMMV. All vendors provide
highly optimized library routines for these.
However, calling those routines results in redundant data
transfers for the involved block vectors.
As discussed in~\cite{Kreutzer15KPM,Pieper15ChebFD} it is possible to fuse all
operations in the $p$-loop of~\Cref{alg:chebfd_vanilla} to a single
algorithm-specific kernel (\textsc{chebfd\_op($H,\vec U, \vec W,\vec
  X$)}) and perform all computations on the three block vectors
(\textsc{$\vec U,\vec W,\vec X$}) once they are in the cache or
register. While this strategy allows for minimum data transfer,
the tailored kernels become very bulky and complex. In particular, for
GPU architectures the resulting CUDA kernel requires manual
architecture-specific tuning as demonstrated in~\cite{Kreutzer15KPM} for
the Nvidia K20.

Another important issue to consider is the storage format of the block
vectors. Here, a row-major approach is beneficial which drives the
irregular access pattern of the SpMV towards streaming access as $n_s$
consecutive block vector elements are loaded for a single matrix
element. Moreover row-major storage also enables SIMD/SIMT
vectorization along the block vector elements and a simple compressed
row storage (CRS) format can be used to store the matrix on all
architectures.

On KNL, the implementation is done via AVX512 compiler intrinsics. The rather
bulky nature of the kernel, together with the use of complex arithmetic,
prevents efficient vectorization of high-level code from the compiler and
necessitates the use of compiler intrinsics.

On the P100 we started with the KPM implementation for the K20m
presented in \cite{Kreutzer15KPM}. Extending this kernel by the update
of the ``filtered'' block vector (line 12 in
\Cref{alg:chebfd_vanilla}) is straightforward but does not change the
computational bottlenecks, which were the reductions required in the
dot products. Here a new feature of the Pascal architecture is
employed: atomic additions using double precision numbers can
increase the performance of the reduction operation for Chebyshev
moments ($\eta_p, \mu_p$).

\subsection{Performance measurement}
\begin{figure}[tpb]
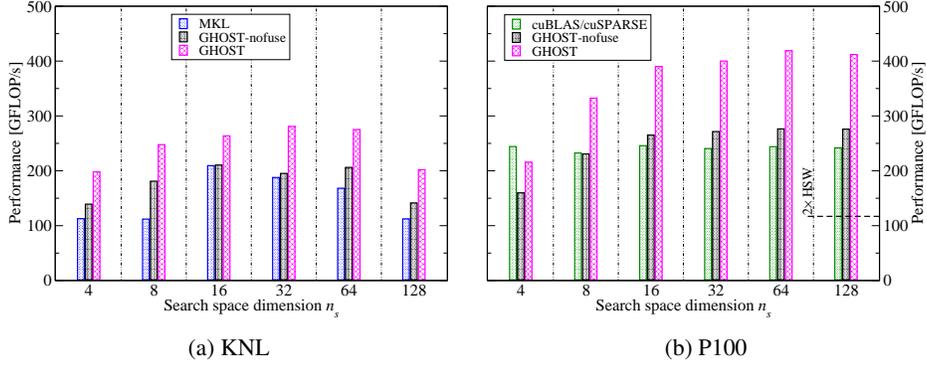

    \centering
    \begin{subfigure}[b]{.48\linewidth}
        \includegraphics*[width=\linewidth]{chebfd+kpm_knl_topi-128-64-64_median.eps}
        \caption{KNL}
        \label{fig:chebfd+kpm_knl_topi}
    \end{subfigure}
    \hfill
    \begin{subfigure}[b]{.48\linewidth}
        \includegraphics*[width=\linewidth]{chebfd+kpm_p100_topi-128-64-64_median.eps}
        \caption{P100}
        \label{fig:chebfd+kpm_p100_topi}
    \end{subfigure}
    \caption{ChebFD performance for the Topi-128-64-64 matrix with $n_p=500$ using different implementations. Maximum performance of a compute node with two Intel Xeon E5-2697v3 processors (Haswell) is shown for reference (data taken from~\cite{Pieper15ChebFD}).}
    \label{fig:chebfd+kpm_topi}
\end{figure}

In \Cref{fig:chebfd+kpm_topi} we present the performance levels which
can be achieved on the KNL and P100 architectures and demonstrate the
need for optimized algorithm-specific kernels.
We find that a tuned implementation of the
\textsc{chebfd\_op()} kernel (labeled ``GHOST'')
outperforms implementations based on a minimum set of standard library calls (SpMMV and BLAS1) typically by 50\%. It is interesting to note that our
manual implementation of those standard calls (``GHOST-nofuse'')
can even outperform the latest vendor-tuned implementations (MKL and
cuBLAS/cuSPARSE). Achieving approximately 10\% of their peak
performance, the two compute devices outperform a standard CPU-based
compute node by a factor of up to four.

\subsection{ChebFD polynomial filter application subspace blocking} \label{sect:subspaceblocking} 

\begin{SCfigure}[1][tb]
\includegraphics*[width=.45\linewidth]{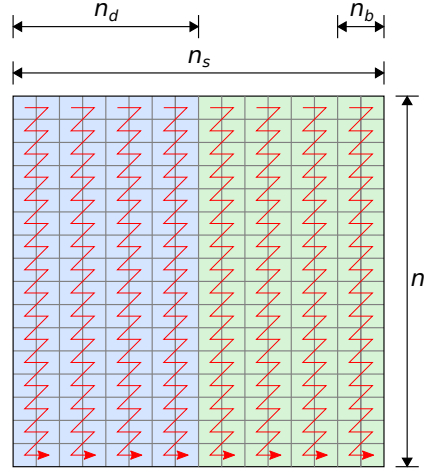}
\caption{Block vector layout for $n_s=16$, choosing $n_b=2$ for filter application
  (subspace) blocking. Pairs of blocks of width $n_b$ will later be used
  for subspace pipelining (see Sect.~\ref{sec:sspipe}).
  The zigzag arrows indicate the storage order of the
  vector elements in memory.}
\label{fig:layout}
\end{SCfigure}

In agreement with published results for CPUs and previous-generation Nvidia
GPGPUs~\cite{Kreutzer15KPM,Pieper15ChebFD} we find that performance
saturates (P100) or even decreases (KNL) at intermediate block vector
size of $n_s=16,32$. To enable optimal performance levels for the
large values of $n_s$ required by the above application scenario, subspace blocking for the block vectors needs to be employed.
Introducing a factor $n_b$ ($ \le n_s$), the application of the filter
can be restricted to a sufficiently small vector block (holding $n_b$
vectors) to achieve optimal performance.
\begin{algorithm}[tb]
\caption{ChebFD polynomial filter application blocking. Here, $n_s$ is assumed to be a multiple of
$n_b$ for simplicity.}
\begin{algorithmic}[1]  
    \For{ $b=0$\ to\ $n_s/n_b-1$ }
    \State $\vec U_b$\ :=\ $\vec u_{bn_b},\dots,u_{(b+1)n_b}$
    \State $\vec W_b$\ :=\ $\vec w_{bn_b},\dots,w_{(b+1)n_b}$
    \State $\vec X_b$\ :=\ $\vec x_{bn_b},\dots,x_{(b+1)n_b}$
        \For{ $n=3$\ to\ $n_p$ }
            \State $\mbox{swap} ( \vec W_b ,  \vec U_b )$
            \State $\Call{chebfd\_op}{H,\vec U_b, \vec W_b, \vec X_b}$
        \EndFor
    \EndFor
\end{algorithmic}
\label{alg:chebfd_blocked}
\end{algorithm}
The corresponding implementation is shown in
\Cref{alg:chebfd_blocked}. Though simple, this code transformation has
strong implications on the row-major data layout of the block
vectors. Row-major ordering now must also be restricted to blocks
containing $n_b$ vectors while the blocks are stored column-wise (see
\Cref{fig:layout}).
We are now free to choose the vector block size independently of
our baseline application, thus we will restrict the following
performance analysis to vector blocks of size $n_b$.

\subsection{Performance analysis}

\label{sec:perfmodel}
We choose the Roofline performance model~\cite{Roofline} to investigate the quality of our implementations and to detect the current hardware bottlenecks:
\begin{equation}
    P^* = \text{min}\left(\Pmax;I(n_b)\times b\right)\;.
\label{eq:roofline}
\end{equation}
This model assumes that the attainable performance is either limited
by in-core execution ($\Pmax$) or by data transfer ($I(n_b)\times b$),
where $b$ is the main memory bandwidth (see \Cref{tab:mem} for typical
values) if data comes from main memory. The arithmetic intensity of
the ChebFD scheme for the Topi test case as a function of $n_b$ is
given by~\cite{Pieper15ChebFD}:
\begin{equation}
    I(n_b) = \frac{146}{260/n_b + 80} \;\frac{\text{Flops}}{\text{Byte}}\; \stackrel{ n_b
      \rightarrow \infty}{\approx} 1.83\;\frac{\text{Flops}}{\text{Byte}}\;.
    \label{intensity}
\end{equation}
This intensity value is calculated as the average numerical
workload and minimum data traffic for applying one matrix row. The
first term in the denominator ($260/n_b \; {\text{Byte}}$) represents
the matrix data traffic: As we have double complex entries,
$20\;\text{Byte}$ per matrix entry (using $4\;\text{Byte}$ indices)
are required. In average one row has 13 entries and we expect to reuse
the row entries for each of the $n_b$ block vector entries. The
minimum data traffic for the three vectors involved in
\textsc{chebfd\_op($H,\vec U, \vec W, \vec X$)} accounts for
$80\;\text{Byte}$ of data traffic, as \textsc{$\vec U$} is read only
(see lines 9-12 in \Cref{alg:chebfd_vanilla}).

Combining Eqs.~\ref{eq:roofline} and~\ref{intensity} it becomes obvious that our
performance measurements are far off the main memory bandwidth limit
of the Roofline model for the full range of $n_b$ in
\Cref{fig:chebfd+kpm_topi} on both architectures. Assuming an
attainable main memory bandwidth of $b=\text{540 GB/s}$ for the P100
($b=\text{470 GB/s}$ for the KNL) the maximum performance should
increase from $P^*=\text{540 GF/s}$ ($P^*=\text{470 GF/s}$) at $n_b=4$
to $P^*=\text{960 GF/s}$ ($P^*=\text{836 GF/s}$) at $n_b=128$ for the
P100 (KNL). This is a first indication that the code is limited on
both architectures not by the available main-memory bandwidth.

Choosing an intermediate value of $n_b=32$, we investigate the actual data transfer volumes
and attained data transfer rates in more detail, see \Cref{tab:perfanalysis}.
\begin{table}
    \centering
    \begin{tabular}{lrrr}
        & Read (GB) & Write(GB) & Bandwidth (GB/s) \\
        \hline
        \hline
        Minimum & 3.77 & 2.15  & - \\
        \hline
        KNL2 MCDRAM & 8.09 & 2.26 & 205.81 \\
        \hline
        P100 HBM2  & 7.02 & 2.28 & 412.17 \\
        P100 L2 & 14.82 & 2.42 & 764.18 \\
        P100 TEX & 38.23 & - & \multirow{2}{*}{\Big\}3129.36} \\
        P100 L1  & 29.96 & 2.42 & 
    \end{tabular}
    \caption{Transferred data volume for memory subsystem
      components and a single ChebFD iteration with the
      Topi-128-64-64 matrix ($n=2.1 \times 10^6$) and $n_b=32$. The
      second row shows minimum data transfers  as assumed in
      the calculation of $I(n_b)$.}
    \label{tab:perfanalysis}
\end{table}
Here, the data volumes have been measured with the Nvidia profiler~\cite{NVprof} and
likwid-perfctr~\cite{Treibig10} for the P100 and the KNL\footnote{Due to the absence
  of suitable tools, measurements were not conducted on the \OFP
  system but on an Intel Xeon Phi 7210 with 64 cores and the same amount
  of L2 cache.}, respectively.

In line with our analysis we find that the actual memory memory
bandwidth rates using MCDRAM and HBM2 are far off the maximum
attainable numbers presented in \Cref{tab:mem}. We also find that the
write data volume matches our assumption underlying
(\ref{intensity}) very well (two vector blocks each of size $n
\times n_s \times 16 \text{ Byte}$ need to be written to main
memory). On the other hand the measured read data volume is
substantially higher than our model assumption indicating that the the
right hand vector block involved in the spMMV is reloaded on the P100
(KNL) approximately four (five) times (see~\cite{Kreutzer2014} for
modeling right hand vector access). As always consecutive chunks of
the block vectors are loaded latency effects are not expected to be
the reason for low memory bandwidth utilization.
\begin{figure}[tpb]
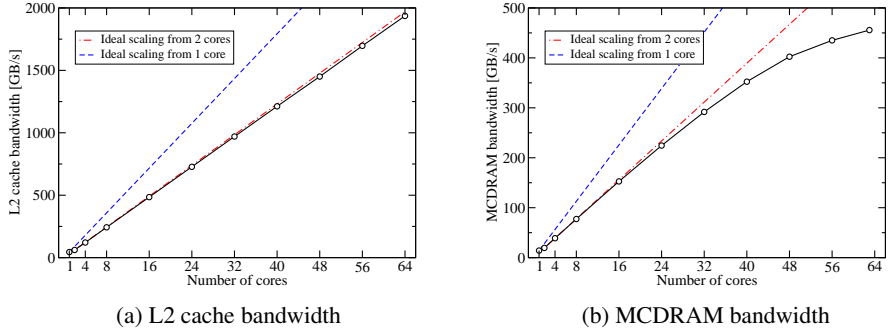

    \centering
    \begin{subfigure}[t]{.48\linewidth}
        \includegraphics*[height=12em]{knightmare_l2_scaling.eps}
        \caption{L2 cache bandwidth}
        \label{fig:knightmare_scaling_l2}
    \end{subfigure}
    \hfill
    \begin{subfigure}[t]{.48\linewidth}
        \includegraphics*[height=12em]{knightmare_mcdram_scaling_smt1.eps}
        \caption{MCDRAM bandwidth}
        \label{fig:knightmare_scaling_mcdram}
    \end{subfigure}
    \caption{Bandwidth scaling of L2 cache and MCDRAM on KNL using
      vector update benchmark explained in the text. L2 measurements were run
      in throughput mode using data sets fitting into the thread-local L2 cache.}
    \label{fig:knightmare_scaling_bandwidth}
\end{figure}

\subsubsection{P100}
The Nvidia profiler identifies the high L1/TEX cache utilization as
primary hardware bottleneck which operates at more than 3 TB/s
bandwidth. Caching right hand side vector elements and warp broadcasts
for reusing the matrix elements across $n_s$ threads may cause this
pressure. This is different from previous results for the K20 presented in
~\cite{Pieper15ChebFD}, where the TEX cache utilization
was also high, but performance was limited by the reduction operations
required in lines 10 and 11 of \Cref{alg:chebfd_vanilla}. This
bottleneck has been removed using the new atomic additions (see
above).
%(KOMMENTAR: HIER KOENNTE MAN ZUMINDEST NOCH L2 READ VOLUME erklaeren: 3.77 GB + 12*1 GB fuer relaoding RHS - aus TEX/L1 sollten dann nochmals min 13 GB TRAFFIC fuer die WARP BCASTs der MAtrixelemente gezogen werden???)
% Kommentar: Deine Shift-taste klemmt schon wieder. Ich würde mir langsam Sorgen machen...

\subsubsection{KNL}
Reliable in-cache data traffic volume measurements were not available
for the KNL architecture at the time of writing. Therefore we
substantiate our expectation that the performance bottleneck is
in-core by a brief scalability analysis of the device architecture and
of our code on one KNL.

\begin{SCfigure}[0.55][tpb]
    \centering
    \includegraphics*[width=.6\linewidth]{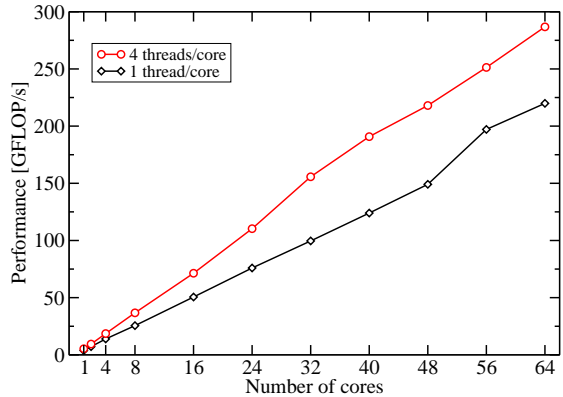}
    \caption{OpenMP scaling of ChebFD performance for Topi-128-64-64
        matrix with $n_p=10/100/500$ for 1-4/8-24/32-64 cores and $n_b=16$ on one KNL.\medskip}
    \label{fig:knl_socketscaling}
\end{SCfigure}

In \Cref{fig:knightmare_scaling_bandwidth} we show attainable
bandwidth values and scalability of MCDRAM and L2 when running a
simple DAXPY kernel. As
the L2 cache segments are shared by two cores and we only perform
local L2 cache accesses we find perfect scalability across the
segments. On contrary, the MCDRAM bandwidth shows the typical
saturation behavior even if only one thread per core is run. However, our
ChebFD implementation scales well across the device and also
benefits substantially from using all SMT threads (see
\Cref{fig:knl_socketscaling}), indicating that neither MCDRAM nor L2
access are the limiting factor. Thus, we identify the in-core execution
of the code to be the bottleneck.\footnote{The identification of the
  respective bottleneck is ongoing but probably pointless as this
  architecture line will not be continued by Intel.}.

In summary, we find that both architectures can achieve approximately
10\% of their theoretical peak performance but are not able to fully
utilize the new memory technology due to in-cache and in-core
bottlenecks despite manually optimized kernels. Hence, future architectural
developments on both systems should focus on improving cache and
core architectures to leverage the additional costs of the new high
bandwidth memory technologies for a broad range of applications.

\subsection{Subspace blocking and large problems}

Often in real world filter diagonalization applications the available
main memory is the limiting factor as one typically aims at large
physical problem sizes ($n$) and a large number of inner eigenstates
($n_s$) at the same time. Thus, the size of the high bandwidth memory
can easily restrict the accessible problem space and may require
(massive) parallelization to provide the required main memory
space. As described in \Cref{sect:Hardware}, the two architectures
under consideration in this work address this problem and allow for
a transparent access to large but slow memory regions located in the
host node (P100) or in a separate DDR4 domain (KNL). We now
investigate the transparent memory access
mechanisms provided on both architectures, i.e., the ``managed mode'' on
P100 and the ``cache mode'' on KNL, to use those large memory spaces
implicitly.

%Extending the subspace blocking strategy presented above towards a
%nested blocking strategy is a straight forward approach to make use of
%the large memory domains for our filter diagonalization application:
%Originally the vectors reside in the host memory and DDR4 domain of
%the P100 and KNL, respectively. If $n_d$ is the maximum number of
%vectors that fits into the 16 GB device memory, we split up the $n_s$
%vectors in blocks of size $n_d$ and run \Cref{alg:chebfd_blocked}
%replacing $n_s$ by $n_d$.
%%layout for the block vectors is sketched.
%%Row-major order storage is
%%chosen for the inner block of $n_b$ consecutive vectors.
%A block of $n_d$ vectors with $n_d$ being a multiple of $n_s$ is
%loaded to the device at same time using the transparent memory access
%mechanisms implicitly. Then, a polynomial filter of degree $n_p$ is
%the applied successively in blocks of $n_s$ vectors to all $n_d$
%device vectors. In \Cref{fig:large_data} the data regions of the full
%block vector which reside on the device memory at the same time have
%the same color. Obviously, $n_b \le n_d \le n_s$.

As we have demonstrated in \Cref{tab:mem}, the transfer rates of
transparent data accesses to the slow memories are very low. In
our case, the time (i.e., the work to be done on the device) between two
accesses to the slow memory is determined by the polynomial filter degree
$n_p$: As shown in \Cref{alg:chebfd_blocked}, a local working set
of $n_b$ vectors is loaded to the high-bandwidth memory and then
reused $n_p-3$ times. Thus the data access to slow memory
may be amortized if  $n_p$ is large enough. 
\begin{SCfigure}[0.55][tp]
    \centering
    \includegraphics*[width=.6\linewidth]{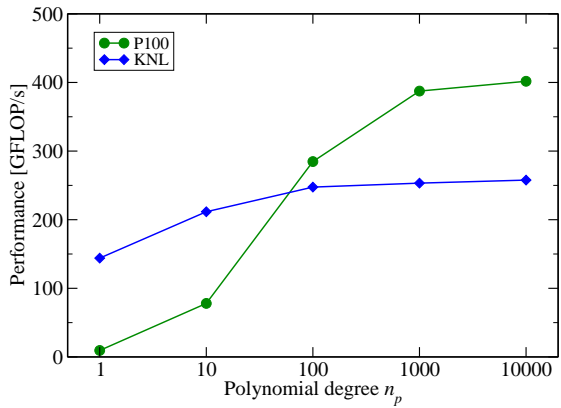}
    \caption{ChebFD performance versus polynomial degree $n_p$
      for the Topi-128-64-64 matrix with
      $n_s=512$ and $n_b=64$ (128) on KNL (P100).\medskip}
    \label{fig:large_data}
\end{SCfigure} 
Indeed we observe no significant impact of the
low-bandwidth memory access for an overall working set of $\approx$60\,\GiB\
beyond $n_p\gtrsim 500$ on P100 and $n_p\gtrsim 100$ on KNL (see \Cref{fig:large_data}).
The different behavior is expected due to the much lower transparent
access bandwidth on the P100 (see \Cref{tab:mem}).

%Thus, overall performance should improve as $n_p$ increases and
%saturate at large $n_p$ , achieving the performance values presented in
%\Cref{fig:chebfd+kpm_topi}. In \Cref{fig:large_data} we have chosen
%the same physical problem as before but have increased the number of
%search vectors to $n_s=512$ ending up in an overall memory footprint
%of more than 60 GB, which is far beyond the capacity of the device
%memory. As expected, performance increases with $n_p$ and saturates at
%the maximum numbers presented above. Of course the P100
%system needs larger $n_p$ to saturate performance as it has a much
%lower transparent access rate and at the same time higher saturated
%performance level.
As discussed in \Cref{sect:setting}, algorithmic efficiency requires
high polynomial degrees $n_p\gg 100$, which matches the demands of
both architectures to achieve high single device performance for large
data sets on the two architectures under consideration.

\section{Large-scale performance} \label{sect:large-scale}
%Experiments on large-scale systems include the Piz Daint) and Oakforest-PACS
%supercomputers. The first is comprised of 5320 computes nodes, each equipped
%with a P100 GPU and an Intel Haswell host CPU (which is ignored in this work).
%Oakforest-PACS, on the other side, contains 8208 nodes of self-hosted KNL
%manycore CPUs. Weak and strong scaling performance is analyzed in the following
%sections.

In this section we present scaling results on both supercomputers
using data sets fitting into the fast memory of both devices.
The \OFP nodes are
operated in ``flat'' mode. Distributed-memory parallelization is done
using the GHOST library~\cite{GHOST}, which supports heterogeneous
parallel execution using an MPI+X approach (currently, X${}\in\{\text{OpenMP},\text{CUDA}\}$).
On Piz Daint we run one MPI
process per host node and on \OFP we use one MPI process and 256
OpenMP threads per KNL node (64 cores). We employ the standard data-level
parallelization approach for ChebFD:
Matrix elements and vector data are distributed across the
MPI processes, each process working with a contiguous set of matrix rows
and the corresponding part of the block vectors. The communication
pattern is determined by the sparsity pattern of the matrix, and
communication of remote block vector elements to local buffers must be
performed before the process-local \textsc{chebfd\_op} is applied
%(i.e., $\vec U_b$; see line 9 in \Cref{alg:chebfd_vanilla}).
% U_b kommt in Alg. 1 gar nicht vor, und so wie er da steht
% auch nicht in Alg. 2
As the matrix structure is reminiscent of a 3D stencil,
nearest-neighbor communication dominates and leads to easy load
balancing and a well-controlled communication volume.

The Cray-proprietary interconnect of Piz Daint uses
a dragonfly network topology. \OFP is based on Intel Omni-Path
with a full fat-tree network built on 48-port
leaf switches and 768-port spine switches. As these networks
should provide sufficient bandwidth for
nearest-neighbor communication, no optimized mapping of MPI ranks to
the topology was done. 
%up to
%intermediate node counts while larger fluctuation hven been seen at
%large node counts.

\subsection{Weak scaling}
\begin{figure}[tpb]
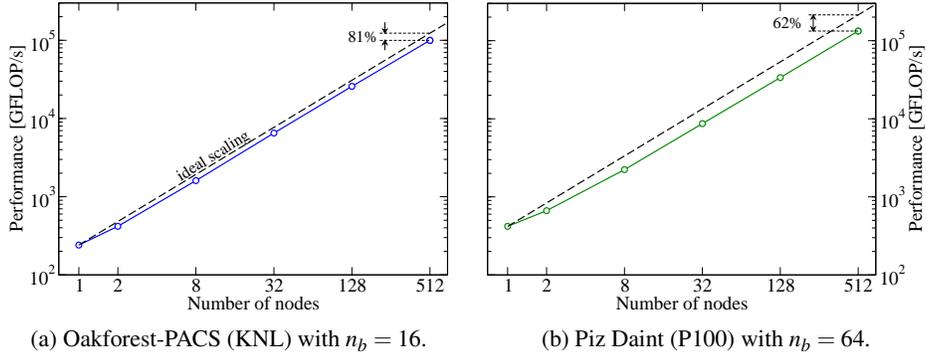

    \centering
    \begin{subfigure}[b]{.48\linewidth}
        \includegraphics*[width=\linewidth]{chebfd+kpm_knl_topi-scaling_vectormode.eps}
        \caption{Oakforest-PACS (KNL) with $n_b=16$.}
        \label{fig:chebfd+kpm_knl_topi_scaling_vectormode}
    \end{subfigure}
    \hfill
    \begin{subfigure}[b]{.48\linewidth}
        \includegraphics*[width=\linewidth]{chebfd+kpm_p100_topi-scaling.eps}
        \caption{Piz Daint (P100) with $n_b=64$.}
        \label{fig:chebfd+kpm_p100_topi_scaling}
    \end{subfigure}
    %\hfill
    %\begin{subfigure}[b]{.48\linewidth}
    %    \includegraphics*[width=\linewidth]{chebfd+kpm_knl_topi-scaling_pipeline.eps}
    %    \caption{Oakforest-PACS (KNL), $n_b=16$, pipelined SpMV with $N_S=128$}
    %    \label{fig:chebfd+kpm_knl_topi_scaling_pipeline}
    %\end{subfigure}
    \caption{Weak scaling of ChebFD in ``vector mode'' with matrices ranging from Topi-128-128-64 (two nodes) to Topi-2048-2048-64 (512 nodes). See the text for details on problem scaling. ($n_s=128$, $n_p=500$)} 
    \label{fig:chebfd+kpm_topi_scaling}
\end{figure}

The weak scaling experiments are based on the problem scaling used
in~\cite{Pieper15ChebFD}. A subdomain of $128 \times 64 \times 64$ is
assigned to each process, corresponding to the Topi-128-64-64 problem
considered so far. For scaling out we
run $2 \times n_\mathrm{scal}^2$ processes ($n_\mathrm{scal}=1,2,4,8,16$) on
a lattice with fixed $z$ dimension that is quadratic in $x$ and $y$, i.e., 
Topi-($128\times n_\mathrm{scal}$)-($64\times 2\times n_\mathrm{scal}$)-$64$ for a given
$n_\mathrm{scal}$.  As long as the communication time is small compared to the
actual computation, which is the case for our choice of parameters, a
simple communication scheme (``vector mode,'' see
\Cref{alg:chebfd_dist_vector}), can be used:
\begin{algorithm}[tb]
    \caption{Blocked application of the ChebFD polynomial filter with explicit and non-overlapping data exchange (``vector mode'').}
\begin{algorithmic}[1]  
    \For{ $b=0$\ to\ $n_s/n_b-1$ }
        \For{ $p=3$\ to\ $n_p$ }
            \State $\mbox{swap} ( \vec W_b ,  \vec U_b )$
            \State \Call{init\_communication}{$\vec U_b$}
            \State \Call{finalize\_communication}{$\vec U_b$}
            \State $\Call{chebfd\_op}{H,\vec U_b, \vec W_b, \vec X_b}$
        \EndFor
    \EndFor
\end{algorithmic}
    \label{alg:chebfd_dist_vector}
\end{algorithm}
Data exchange using non-blocking MPI
(lines 4 and 5) is separated from the
process-local computation ($\Call{chebfd\_op}{H,\vec U_b, \vec W_b,
  \vec X_b}$). The weak scaling performance results for both systems
are shown in \Cref{fig:chebfd+kpm_topi_scaling} for up to 512 nodes. 
The communication overhead introduced by the vector mode is visible
for two and eight nodes because communication sets in first in the $y$
direction at $n_\mathrm{scal}=1$ and then additionally in the $x$
direction at $n_\mathrm{scal}=2$.  From eight nodes onward we see
perfect scaling since per-node communication and computation times
stay constant. Based on the single-node performance the systems
achieve a parallel efficiency of 81\% (\OFP) and 62\% (Piz Daint),
which compares to 73\% obtained on the SuperMUC-Phase2
system~\cite{Pieper15ChebFD}.
The lower efficiency of Piz Daint can be attributed to the transfer of
vector data over the PCIe bus (even though our implementation uses
GPUdirect communication avoiding an intermediate copy of communication
data in the host memory) and the high single-node performance. Still
Piz Daint provides best absolute performance, achieving 132.7\,TF/s
at 512 nodes, thereby outperforming the CPU-only SuperMUC-Phase2
results in~\cite{Pieper15ChebFD} by a factor of three at the same node
count. The underlying numerical problem considered here is the
computation of approximately 100 inner eigenvalues $ (\le n_s)$ of a
matrix of dimension $n=10^9$.

\subsection{Strong scaling and subspace pipelining}\label{sec:sspipe}

Sparse linear algebra problems often show
limited strong scalability because the computation time per
process decreases faster than the corresponding communication
time. Our vector mode implementation, which was acceptable with
weak scaling and rather large per-node problem sizes, must thus
be improved by explicitly overlapping communication with
computation. A typical approach to this problem is
to do local computations (i.e., handle matrix elements which only
access local vector elements) while communicating the non-local vector
elements and then doing the remaining work with the just-received data,
updating the partial results~\cite{Kreutzer2012}. This implementation needs
to update the local result vector twice and thus increases the
main memory data traffic. Modifying the subspace blocking scheme
introduced in \Cref{sect:subspaceblocking} towards pipelining of
computation and communication steps of successive filter applications
as presented in \Cref{alg:chebfd_dist_pipe} offers an interesting
alternative.
\begin{algorithm}[tb]
    \caption{Blocked application of the ChebFD polynomial filter with pipelined communication and computation (``pipelined mode'').}
    \begin{algorithmic}[1]  
        \For{ $p=3$\ to\ $n_p$ }
            \State $\mbox{swap} ( \vec W ,  \vec U )$
            \State \Call{init\_communication}{$\vec U_0$}
            \State \Call{finalize\_communication}{$\vec U_0$}
            \For{ $b=0$\ to\ $n_s/n_b-2$ }
                \State \Call{init\_communication}{$\vec U_{b+1}$}
                \State $\Call{chebfd\_op}{H,\vec U_b, \vec W_b, \vec X_b}$
                \State \Call{finalize\_communication}{$\vec U_{b+1}$}
            \EndFor
            \State $\Call{chebfd\_op}{H,\vec U_{n_s/n_b-1}, \vec W_{n_s/n_b-1}, \vec X_{n_s/n_b-1}}$
        \EndFor
    \end{algorithmic}
    \label{alg:chebfd_dist_pipe}
\end{algorithm}

\begin{SCfigure}[0.55][tpb]
    \centering
    \includegraphics*[width=.6\linewidth]{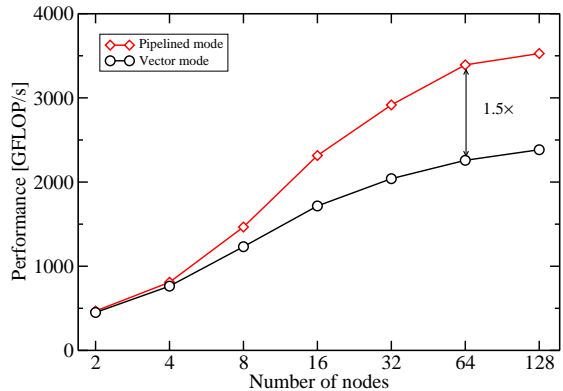}
    \caption{Strong scaling ChebFD performance for the Topi-128-128-64
      matrix with $n_s=128$, $n_p=500$ and $n_b=16$ on \OFP,
      comparing pipelined and vector communication modes.\medskip}
    \label{fig:strongscaling}
\end{SCfigure}

Instead of calculating the full polynomial for a given block of $n_s$
vectors, the polynomial degree for the full block vector is increased
step by step. The inner loop runs over the full block vector and
the computation on the current subblock can be overlapped with the
communication required for the next
subblock. This strategy avoids the overhead of
writing the result vector twice, maintaining the same computational
intensity as the non-MPI code. As long as $n_b/n_s$ is
sufficiently large and asynchronous communication is supported by the
MPI implementation, the communication should be effectively hidden.
A comparison of vector mode and subspace pipelining for
strong scaling on \OFP is given in \Cref{fig:strongscaling}
for the Topi-128-128-64 problem ($n=4 \times 10^6$).
As expected, the benefit of subspace pipelining increases as the number
of processes goes up because the communication becomes more relevant.
A maximum speed-up with respect to vector mode of 50\% could be observed in
this test. Note that the
speed-up of communication-hiding approaches in SpMV is limited to a factor of
two. Further increasing the processor count will diminish the
benefit of subspace pipelining as we reach the completely
communication-bound regime.
%
%
%As opposed to weak scaling, strong scaling includes a constant problems size for an
%increasing number of processes. In this scenario, the computation
%effort per processes decreses for an increasing process count, while
%the overall communication effort increases. Obviously, this
%contradicts the requirements for efficient ``vector mode'' computation
%%as described above. For algorithms containing large subspaces (like
%the present ChebFD), subspace pipelining can be employed to achieve an
%efficient overlap of computation and communication.
%\Cref{alg:chebfd_dist_pipe} illustrates this idea: First, an initial
%sub-block of the input vector is communicated (``wind-up phase''). The
%according result subspace is then computed and at the same time, the
%next sub-block is communicated. Finally, the last sub-block is
%computed (``wind-down phase''). An important tuning factor for this
%implementation is the innermost block vector width $n_b$. It should be
%large enough to allow for an efficient \textsc{textsc\_op} (see above
%for the influence of $n_b$ on the performance), but small enough to
%minimize the effect of the wind-up and wind-down phases.
%

On Piz Daint subspace pipelining did not show any benefits. With
low-level experiments we have checked that non-blocking MPI
communication using GPUdirect does not overlap with GPU
computation (Cray is currently investigating this problem).
%\begin{itemize}
%    \item Subspace pipelining for efficient and explicit
%        communication-computation overlap
%    \item shown in \cref{alg:chebfd_dist_pipe}
%    \item ineffient on Piz Daint: GPUdirect communication cannot
%        overlap with GPU kernels
%    \item best benefit for communication-intensive scenarios
%    \item strong scaling performance on \cref{fig:strongscaling}
%    \item the higher the communication effort, the more one can profit from
%        pipelining
%    \item trade-off $n_b$: large enough to give optimal performance (and save on
%        matrix transfers), small enough to have insignificant wind-up and
%        wind-down phase
%\end{itemize}

\section*{Summary}
This work has investigated performance properties and subspace blocking optimization techniques for a Chebyshev filter diagonalization (ChebFD) algorithm on the  Intel Xeon Phi (``Knights Landing'') and Nvidia P100 (``Pascal'') architectures. Our block vector implementation achieves approximately 10\% of the theoretical peak performance and is no longer memory bound on both architectures. We have demonstrated that subspace blocking with a sufficiently large polynomial filter degree enables efficient use of the complete node-level address space (i.e., high-bandwidth and low-bandwidth memory) transparently without impacting the node performance even if the working set exceeds the high-bandwidth memory size by far. Subspace blocking can be further extended towards a pipelining of the communication and computation phase in the filter application, which allows for simple communication hiding. Though this study has focused on using ChebFD in the context of inner eigenvalue computations for topological insulators, the basic strategies presented can be applied to many applications evaluating Chebyshev polynomials of large sparse matrices and should be of interest for block formulations of iterative solvers in sparse linear algebra.
%The GHOST software library used in this work is available for download~\cite{GHOSTSW}.

\section*{Acknowledgments} 

%\GHcomment{Hidden for double-blind review}
This work was funded by DFG SPP1648 through the ESSEX-II project and
by a grant from the Swiss National Supercomputing Centre (CSCS) under
project ID d35. We gratefully acknowledge the access to the \OFP
supercomputers at JCAHPC, Univ. of Tokyo.
HF and GW gratefully
acknowledge the hospitality of Los Alamos National Laboratory.

\bibliographystyle{splncs03}
\bibliography{lspp17}

\end{document}